\documentclass{article}
\usepackage{graphicx}
\usepackage{dcolumn}
\usepackage{bm}
\usepackage{amssymb}
\usepackage{amsmath}
\usepackage{color}

\newcommand\br{\begin{eqnarray}}
\newcommand\er{\end{eqnarray}}
\newcommand\be{\begin{equation}}
\newcommand\ee{\end{equation}}

\newcommand\bc{\begin{center}}
\newcommand\ec{\end{center}}

\newcommand{\nn}{\nonumber \\}

\newcommand\n{\nu}

\newcommand{\twopartdef}[4]
{
	\left\{
		\begin{array}{ll}
			#1 & \mbox{if } #2 \\
			#3 & \mbox{if } #4
		\end{array}
	\right.
}

\begin{document}
\title{General Rule for Boundary conditions from the Action Principle}

\author
{Roee Steiner \footnote{e-mail: roeexs@gmail.com} \\ Physics Department, Ben Gurion University of the Negev
\\ Beer Sheva 84105, Israel}
\date{\today}

\maketitle
\begin{abstract}
We construct models where initial and boundary conditions can be found from the fundamental rules of physics, without the need to assume them, they will be derived from the action principle. Those constraints are established from physical view point, and it is not in the form of Lagrange multipliers. We show some examples from the past and some new examples which can be useful, where constraint can be obtained from the action principle. Those actions represent physical models. We show that it is possible to use our rule to get those constraints directly. 
\end{abstract}

\section{Introduction}
In the last few years we have worked on the question of boundary condition from the action principle.
We have found some different ways to establish boundary condition from the action, and which we motive it, to answer the question of the initial condition for the inflation theory, or to the question of confinement.
In physics we deal with equations of motion that are obtained by varying the action with respect to different fields, here the question of the initial condition or boundary condition are normally separated from the equation of motion, and by giving them both we can solve the physical problem (like in many differential equation problems where the solution is determined by the initial condition).
Knowing just the equation of motion or just the initial conditions does not give the solution of the problem. 
Landau said " The future physical theory should contain not only the basic
equations but also the initial conditions for them " [L.D. Landau according to I.M. Khalatnikov].
From this point we are motivated to construct a model where initial conditions can be found from the fundamental rules of physics, without the need to assume them, they will be derived.\\
In section II we derive a general rule which one can use to find or establish constraint from the action principle. Those constraints are established from a physical view point, and it is not of the form of Lagrange multiplier. In section III we show some special cases in which the rule can not be established. In section IV we shows four examples from the past in which produce constraints from the action principle. We show that it is possible to use our rule to get those constraint directly. 
We give two appendixes, one concerning with the definition of charge in terms of a dynamical field. The second uses non abelian charge. 
 
\section{General rule for constraint from the action principle}
If we have an action in the general form:
 \begin{equation}\label{eq:action with + g function 2}
S=\int d^{4}x \{\mathcal{L} + \mathcal{G}(g (f))\}
\end{equation}
Where $ \mathcal{L} $ and $ \mathcal{G} $ are invariant under gauge transformation.
We require that there is not exist any transformation or field redefinition which $ \mathcal{L} + \mathcal{G} \Rightarrow \mathcal{L} $.\\The function $ g(f(x)) $  have singular derivatives on some surface $ f(x)= constant $ (we include the case of step function as singular derivative situation), where $ f(x) $ is some analytic function. Because $ g(f(x)) $ will be discountenances on the surface $ f(x) = const $ then the equation of motion will have the constraint equation: 
\begin{eqnarray}\label{eq general constraint}
\frac{\partial^{2} \mathcal{G}}{\partial (\partial_{\mu}\phi)\partial g(x)}\partial_{\mu}f(x)\mid _{x\in f(x) = const} = 0
\end{eqnarray}
on the surface $ f(x) = const $ where $ g(x) $ have singular derivative. The $ \delta \partial \phi $ is a variation of the derivative of the field which can be a scalar , Dirac or vector.\\
\textbf{Proof:} \\
Lets define $ \mathcal{G}=\mathcal{G}(\phi,\partial_{\mu}\phi , g(f(x))) $ where $ g(f(x)) $ is some function that is gauge invariant but have singular derivative in some surface $ f(x) = const $. We will derive Euler Lagrange on the action \ref{eq:action with + g function 2}:
\begin{eqnarray}
& \frac{\partial\mathcal{L}}{\partial \phi} - \partial_{\mu}\frac{\partial \mathcal{L}}{\partial (\partial_{\mu}\phi)} 
+ \frac{\partial\mathcal{G}}{\partial \phi} - \partial_{\mu}\frac{\partial \mathcal{G}}{\partial (\partial_{\mu}\phi)} = 
\nn & \frac{\partial\mathcal{L}}{\partial \phi} - \partial_{\mu}\frac{\partial \mathcal{L}}{\partial (\partial_{\mu}\phi)} 
+ \frac{\partial\mathcal{G}}{\partial \phi} 
\nn & - \frac{\partial^{2} \mathcal{G}}{(\partial (\partial_{\mu}\phi))^{2}}\partial_{\mu}\partial^{\mu}\phi - \frac{\partial^{2} \mathcal{G}}{\partial (\partial_{\mu}\phi)\partial \phi}\partial_{\mu}\phi 
\nn & - \frac{\partial^{2} \mathcal{G}}{\partial (\partial_{\mu}\phi)\partial g(x)}\frac{\partial g(f(x))}{\partial f(x)}\partial_{\mu}f(x) = 0
\end{eqnarray} 
The term $ \frac{\partial g(f(x))}{\partial f(x)} $ is singular on the surface $ f(x) = const $ so we must conclude that 
\begin{eqnarray}
\frac{\partial^{2} \mathcal{G}}{\partial (\partial_{\mu}\phi)\partial g(x)}\partial_{\mu}f(x)\mid _{x\in f(x) = const} = 0
\end{eqnarray}
\section{Special Cases}
\subsection{Transformation that kill undefined term}
If there is some transformation which eliminate the $ \mathcal{G} $ term so that, 
\begin{equation}
\mathcal{L} + \mathcal{G} \Rightarrow \mathcal{L} 
\end{equation}
then $ \mathcal{G} $ will not produce any constraint on the system.\\
Further more, the solution of the equation of motion of the action \ref{eq:action with + g function 2} is the solution of the equation of motion of the action $ \int d^{4}x (\mathcal{L}) $ with the transformation which produce $ \mathcal{L} \Rightarrow \mathcal{L} - \mathcal{G} $.\\
A nice example which was widely investigated in ref \cite{GuendelmanRoee} is the case where $ \mathcal{L} $ is the ordinary Dirac Lagrangian
\begin{eqnarray}
\mathcal{L} = \bar{\psi}(\frac{i}{2}\gamma^{\alpha}\stackrel{\leftrightarrow}{\partial}_{\alpha}-m)\psi
\end{eqnarray}
and where the function $ \mathcal{G} $ which have a singular derivative is:
\begin{align}
& \mathcal{G} = \int (\bar{\psi}\gamma^{0}\psi)\, \delta(t-t_{0})\, d^{3}x
\end{align}
The action $ \int (\mathcal{L} + \mathcal{G}) d^{4}x $ illustrate physical model where the global charge in the universe are part of the local fermion system, which is Mach like principle for global charge.
\\If the solution of $ \mathcal{L} $ is $ \psi_{D} $, then the transformation $ \psi = e^{i \int(\bar{\psi}\gamma^{0}\psi ) \, \theta(t-t_{0})d^{3}x}\psi_{D} $ transform the solution of $ \mathcal{L} + \mathcal{G} $ into the solution of $ \mathcal{L} $.
\subsection{$ f(x) $ as dynamical independent field}
There is a special case when:
\begin{eqnarray}
S=\int d^{4}x (\mathcal{L}_{\phi} + \mathcal{L}_{f} + \mathcal{G}(g(f(x))))
\end{eqnarray}
where:
\begin{eqnarray}
\mathcal{G} = J^{\mu}(\phi , \partial^{\mu}\phi)\,g(f(x))\, \partial_{\mu}f(x)
\end{eqnarray}
$ J^{\mu}(\phi , \partial^{\mu}\phi) $ is some vector that depends on $ \phi $ and $ \partial^{\mu}\phi $, and $ g(f(x)) $ is some function of dynamical field $ f(x) $, where $ g(f(x)) $ has undefined derivative on the surface $ f(x) = const $.
It is easy to see that equation \ref{eq general constraint} still gives constraint on the field $ \phi $, but if we derive Euler Lagrange on the field $ f(x) $ then we get:
\begin{eqnarray}
\frac{\partial\mathcal{L}_{f}}{\partial f} - \partial_{\mu}\frac{\partial \mathcal{L}_{f}}{\partial (\partial_{\mu}f)} - \partial_{\mu}J^{\mu}(\phi , \partial^{\mu}\phi) g(f) = 0
\end{eqnarray} 
which does not have any undefined point, so we don't have constraint on the field $ f(x) $.
Furthermore we can see that if $ J^{\mu}(\phi , \partial^{\mu}\phi) $ is a conserved current term or equivalently $ \partial_{\mu}J^{\mu}(\phi , \partial^{\mu}\phi)=0 $ then the field $ f(x) $ can be independent field (it depends on the potential).
\section{Example of models where constraint cannot be avoided}
In this section we will show some models that give constraint. We will see that the constraint of those model can be follow immediately from equation \ref{eq general constraint} which is general for constraint model of this form. We should emphasize that the action of those examples have been built around some physical philosophy, and the equation of motion produce the physical constraints. The MIT bag model have been built around the idea that the quarks are free to move in some cavity. The next examples show physical system where the global charge in the universe are part of the local scalar field system, which is Mach like principle for global charge. 
\subsection{MIT bag model} 
In the M.I.T bag model \cite{mitbag} (for review see \cite{mitbag1}) they produce a model that can give confinement mechanism. Following action has studied \cite{Chodos}:
\begin{eqnarray}\label{mit action}
S= \int_{V} d^{4}x[\, \bar{\psi}(\frac{i}{2}\gamma^{\mu}\stackrel{\leftrightarrow}{\partial}_{\mu} -m)\psi +\partial_{\mu}(\lambda^{\mu}\bar{\psi}\psi) - B ]
\end{eqnarray}
where the integration is under the volume of the bag, $ B $ is some constant and $ \lambda^{\mu} $ is some vector. From this action it is follow that the equation of motion inside the bag (the volume) is the Dirac equation, and out side the bag is zero. On the surface they got the constraint equation:
\begin{eqnarray}\label{mit surface constraint}
\frac{i}{2}n_{\mu}\gamma^{\mu}\psi + n_{\mu}\lambda ^{\mu} \psi = 0 
\end{eqnarray}
Which from the knowledge that $ (in_{\mu}\gamma^{\mu})^{2} = 1 $ one gets (squaring equation \ref{mit surface constraint}) that:
\begin{align}
4(n_{\mu}\lambda^{\mu})^{2} = n^{2}
\end{align}
and 
\begin{align}
\bar{\psi}\psi = 0
\end{align}
on the surface.\\
The action \ref{mit action} can be seen as:
\begin{eqnarray}\label{mit action2}
 \int d^{4}x \,\, \Theta(f(x)) [ \bar{\psi}(\frac{i}{2}\gamma^{\mu}\stackrel{\leftrightarrow}{\partial}_{\mu} -m)\psi +\partial_{\mu}(\lambda^{\mu}\bar{\psi}\psi) - B ]
\end{eqnarray}
where $ \Theta $ is step function and $ f(x) $ is some function that define the volume of the bag.
It is easy to see that equation \ref{mit surface constraint} which define the surface constraint of the M.I.T bag model can be follow easily by equation \ref{eq general constraint} of our theory, where
\begin{align}
\mathcal{G} = \Theta(f(x)) [ \bar{\psi}(\frac{i}{2}\gamma^{\mu}\stackrel{\leftrightarrow}{\partial}_{\mu} -m)\psi +\partial_{\mu}(\lambda^{\mu}\bar{\psi}\psi) - B ]
\end{align}
 and 
\begin{align}
g(f(x)) = \Theta(f(x))
\end{align} 
so:
\begin{eqnarray}
& \frac{\partial^{2} \mathcal{G}}{\partial (\partial_{\mu}\bar{\psi})\partial \Theta(f(x))}\partial_{\mu}f(x)\mid _{x\in f(x) = const} =
 \nn & \frac{i}{2}(\partial_{\mu}f)\gamma^{\mu}\psi + (\partial_{\mu}f(x))\lambda ^{\mu} \psi = 0
\end{eqnarray} 
which is exactly equation \ref{mit surface constraint} where $ n_{\mu} = \partial_{\mu}f(x) $.

\subsection{Initial condition from action with general potentials depending on charge}
We review example of constraint that we have got in our paper \cite{GuendelmanRoee2}, in which we have constraint on the scalar field to be zero on some time surface, we will see that this constraint can be found easily by equation \ref{eq general constraint} of the new theory. We will begin with the action of Klein Gordon equation (with the metric $ diag(-1,1,1,1) $):
\begin{eqnarray}\label{general potential: action 2}
&S=\int d^{4}x\sqrt{-g}\,[(\partial^{\mu}\phi^{*} +i\frac{g'}{2} A^{\mu}\phi^{*})(\partial_{\mu}\phi -i\frac{g'}{2} A_{\mu}\phi)
\nn & -V(\phi,\phi^{*},Q)]
\nonumber\\& -\frac{1}{4}\int F^{\mu\nu}F_{\mu\nu}\sqrt{-g} d^{4}x - \frac{1}{16\pi G}\int{\sqrt{-g}R\,d^{4}x}=
\nonumber\\ & \int{d^{4}x\sqrt{-g}\,[(D\phi)^{*}(D\phi)-V(\phi,\phi^{*},Q)]}
\nn & -\frac{1}{4}\int F^{\mu\nu}F_{\mu\nu}\sqrt{-g}  d^{4}x - \frac{1}{16\pi G}\int{\sqrt{-g}R\,d^{4}x}
\end{eqnarray}
where the $ Q $ that appears in the potential $ V $ is given by:
\begin{eqnarray}\label{eq: def m KG}
& Q=\lambda\int{d^{3}y\,\sqrt{-g}[\phi^{*}i\stackrel{\leftrightarrow}{\partial^{0}}\phi + g'A^{0}\phi^{*}\phi]}\mid_{y^{0}=t^{0}}=
\nn & \lambda\int{d^{4}y\,\sqrt{-g}[\phi^{*}i\stackrel{\leftrightarrow}{\partial^{0}}\phi + g'A^{0}\phi^{*}\phi]\delta(y^{0}-t_{0})}
\end{eqnarray}
which is the total charge in the universe by the definition of Klein Gordon field.
So by variation we get the equation of motion:
\begin{eqnarray}\label{eq:equation motion Klein Gordon 5435}
&-\partial_{\mu}(\sqrt{-g}g^{\mu\nu}\partial_{\nu}\phi) - i \frac{g'}{2}\partial_{\mu}(\sqrt{-g} A^{\mu}\phi)
\nn & - i\sqrt{-g}\frac{g'}{2} A_{\mu}\partial^{\mu}\phi +\sqrt{-g} (\frac{g'}{2})^{2}A_{\mu}A^{\mu}\phi
 -\sqrt{-g}\frac{\partial V}{\partial \phi^{*}} 
\nonumber\\& -2i\sqrt{-g}\lambda(\int{d^{4}x\,\sqrt{-g}\,\frac{\partial V}{\partial Q}})\delta(y^{0}-t_{0})[\partial^{0}\phi -i\frac{g'}{2}A^{0}\phi]
\nn & -\lambda(\int{d^{4}x\,\sqrt{-g}\,\frac{\partial V}{\partial Q}})i\phi \,\partial^{0}(\sqrt{-g}\delta(y^{0}-t_{0}))=0
\end{eqnarray}

if we do the transformation 

\begin{equation}\label{eq:Klein A transform}
A^{0}\longrightarrow A^{0}+\frac{2i\lambda_{1}b}{g'}\delta(y^{0}-t_{0})
\end{equation}
and
\begin{equation}\label{eq:Klein psi transform}
\phi=e^{\lambda_{2} b\theta(y^{0}-t_{0})}\phi_{0}
\end{equation}
where $ b=i\lambda(\int{d^{4}x\,\sqrt{-g}\,\frac{\partial V}{\partial Q}}) $
\\we have that:

\begin{eqnarray}\label{eq:equation motion Klein Gordon 2}
&-\partial_{\mu}(\sqrt{-g}g^{\nu\mu}\partial_{\nu}\phi_{0}) - i \frac{g'}{2}\partial_{\mu}(\sqrt{-g}A^{\mu}\phi_{0})
\nn & - i\sqrt{-g}\frac{g'}{2} A_{\mu}\partial^{\mu}\phi_{0} + \sqrt{-g}(\frac{g'}{2})^{2}A_{\mu}A^{\mu}\phi_{0}-\sqrt{-g}\frac{\partial V}{\partial \phi^{*}} \nonumber\\&
-2b\sqrt{-g}\delta(y^{0}-t_{0})[(\lambda_{1}-\lambda_{2}+1)(\partial^{0}\phi_{0}-i\frac{g'}{2} A^{0}\phi_{0})\n & 
\nn & +\frac{1}{2} b\delta(y^{0}-t_{0})\phi_{0}(-\lambda^{2}_{2}+(2\lambda_{1}-\lambda^{2}_{1})+2(\lambda_{1}-1)\lambda_{2})]\nonumber\\&-b(\lambda_{1}-\lambda_{2}+1)\phi_{0}\,\partial^{0}(\sqrt{-g}\delta(y^{0}-t_{0})) = 0
\end{eqnarray}
if we require that the equation (\ref{eq:equation motion Klein Gordon 2}) will be like the ordinary Klein Gordon equation where there are no delta function appear, since those delta functions represent singular interactions, we need that:

\begin{eqnarray}
\lambda_{1}-\lambda_{2}+1=0\\
-\lambda^{2}_{2}+(2\lambda_{1}-\lambda^{2}_{1})+2(\lambda_{1}-1)\lambda_{2}=0
\end{eqnarray}

But there is no solution for $ \lambda_{1} $ and $ \lambda_{2} $ for those two equation.
If we will say that the covariant derivative is equal to zero $ \partial^{0}\phi_{0}-i\frac{g'}{2} A^{0}\phi_{0}=0 $ and $ \lambda_{1}-\lambda_{2}=2 $ then we still have problem with the term $ \partial^{0}\delta(y^{0}-t_{0}) $ in equation \ref{eq:equation motion Klein Gordon 2}. So we must to say that:
\begin{equation}
\phi^{*}(t=t_{0})\phi(t=t_{0})=0
\end{equation}
where $ \lambda_{1}-\lambda_{2}+1=0 $ which eliminates all the delta term in equation \ref{eq:equation motion Klein Gordon 2}.\\
We can see that the same result can be found by using equation \ref{eq general constraint} by setting $ \mathcal{G} = V(\phi,\phi^{*}, Q) $ and $ g(f(x))= \delta(y_{0}- t_{0}) $ so:
\begin{eqnarray}
& \frac{\partial^{2} \mathcal{G}}{\partial (\partial_{\mu}\phi)\partial g(x)}\partial_{\mu}f(x)\mid _{x\in f(x) = const} =
\nn & \frac{\partial^{2} \mathcal{G}}{\partial (\partial_{\mu}\phi ^{*})\partial \delta(y_{0}-t_{0})}\mid _{t_{0}} = \lambda(\int{d^{4}x\,\sqrt{-g}\,\frac{\partial V}{\partial Q}})\phi(t_{0}) = 0
\end{eqnarray}
which gives $ \phi(t_{0}) = 0 $.
This model can be used for creating initial condition for inflation.\\
Different constraint can be follow by different charge definition. In the paper \cite{GuendelmanRoee2}, we showed that also boundary condition can be contract by define the charge in different hyper- surface and which is not time like surface. 

\subsection{Constraint controlled by a dynamical field}
We now represent an action of two scalar field , where the potential is $ V(\phi,\phi^{*},f,Q) $. The $ Q $ term is defined in equation \ref{Theta} (see appendix):
\begin{eqnarray}\label{boundery: action}
&S=\int d\sigma\sqrt{-g}\,[(\partial_{\mu}\phi^{*} +i\frac{g'}{2} A_{\mu}\phi^{*})(\partial^{\mu}\phi -i\frac{g'}{2} A^{\mu}\phi)
\nn & + \partial_{\mu}f \partial^{\mu}f -V(\phi,\phi^{*},f,Q) ]
\nonumber\\&-\frac{1}{4}\int F^{\mu\nu}F_{\mu\nu}\sqrt{-g} d\sigma  - \frac{1}{16\pi G}\int{\sqrt{-g}R\,d\sigma}=
\nonumber\\&\int{d\sigma\sqrt{-g}\,[(D_{\mu}\phi)^{*}(D^{\mu}\phi)+ \partial_{\mu}f \partial^{\mu}f -V(\phi,\phi^{*},f,Q)]}
\nn & -\frac{1}{4}\int F^{\mu\nu}F_{\mu\nu}\sqrt{-g}  d\sigma - \frac{1}{16\pi G}\int{\sqrt{-g}R\,d\sigma}
\end{eqnarray}
from this action by variation on $ \phi^{*} $ , we get the equation of motion:
\begin{align}\label{eq:equation motion Klein Gordon 543}
&-\partial_{\mu}(\sqrt{-g}g^{\mu\nu}\partial^{\nu}\phi) - i \frac{g'}{2}\partial_{\mu}(\sqrt{-g}A^{\mu}\phi)
\nn & - i\sqrt{-g}\frac{g'}{2} A_{\mu}\partial^{\mu}\phi
 + \sqrt{-g}(\frac{g'}{2})^{2}A_{\mu}A^{\mu}\phi -\sqrt{-g}\frac{\partial V}{\partial \phi^{*}} 
\nonumber\\& -\sqrt{-g}(\int{d\sigma\,\sqrt{-g}\,\frac{\partial V}{\partial Q}})\delta (f(x) - f_{0})\partial^{\mu}f(x)[2i\partial_{\mu}\phi-g'A_{\mu}\phi]
\nn &
-(\int{d\sigma\,\sqrt{-g}\,\frac{\partial V}{\partial Q}})i\phi \,\partial_{\mu}(\sqrt{-g}\delta (f(x) - f_{0})\partial^{\mu}f(x))=0
\end{align}
if we do the transformation 
\begin{equation}\label{eq:Klein A transform mu}
A^{\mu}\longrightarrow A^{\mu}+\frac{2i\lambda_{1}b}{g'}\delta (f(x) - f_{0})\partial^{\mu}f(x)
\end{equation}
and
\begin{equation}\label{eq:Klein psi transform mu}
\phi=e^{\lambda_{2} b\theta(f(x) - f_{0})}\phi_{0}
\end{equation}
where $ b=i(\int{d\sigma\,\sqrt{-g}\,\frac{\partial V}{\partial Q}}) $
\\we have that there is no solution for $ \lambda_{1} $ and $ \lambda_{2} $, so we must conclude that
\begin{eqnarray}\label{constraint dynamical}
\phi(x_{0})\partial_{\mu}f(x_{0})\partial^{\mu}f(x_{0})=0
\end{eqnarray} 
when $ f(x_{0})=f_{0} $
\\The equation of motion of the scalar field $ f $ did not influenced by the $ Q $ term and it is:
\begin{eqnarray}
\partial_{\mu}(\sqrt{-g}g^{\mu\nu}\partial_{\nu}f) + \sqrt{-g}\frac{\partial V}{\partial f} = 0
\end{eqnarray}
It is easy to see that the same constraint (equation \ref{constraint dynamical}) can be found directly from equation \ref{eq general constraint}
\subsection{Initial condition for Non Abelian fields}
One can define charge for non Abelian field which is constant and gauge invariant.
This definition have been used in the past and it is a private case of the Abbott and Deser definition \cite{ABBOTT} of non abelian charge in the case of zero background field.\\ 
For convocation we going to show the derivation of the charge definition .
As we know in non Abelian $ SU(N)$ the definition of $ F^{\mu\mu \, a} $ is:
\begin{eqnarray}\label{F def}
F^{\mu\nu  a} = \partial^{\mu}A^{\nu  a} - \partial^{\nu}A^{\mu  a} + g \epsilon^{abc}A^{\mu b}A^{\nu c}
\end{eqnarray}
The coveriant derivative is:
\begin{eqnarray}\label{DF def}
D_{\mu}F^{\mu\nu a} = \partial_{\mu}F^{\mu\nu a} + g\epsilon^{abc}A^{b}_{\mu}F^{\mu\nu c} = J^{\nu a}
\end{eqnarray}
where $ J^{\nu a} $ is the current, which depends on scalar or Dirac fields.
By entering \ref{F def} to \ref{DF def} we get:
\begin{eqnarray}\label{F in DF}
& \partial_{\mu}(\partial^{\mu}A^{\nu  a} - \partial^{\nu}A^{\mu  a}) + g \epsilon^{abc}\partial_{\mu}(A^{\mu b}A^{\nu c}) \nn & + g\epsilon^{abc}A^{b}_{\mu}(\partial^{\mu}A^{\nu  c} - \partial^{\nu}A^{\mu  c})
\nn & + g^{2}\epsilon^{abc}A^{b}_{\mu} \epsilon^{cdf}A^{\mu d}A^{\nu f} = J^{\nu a}
\end{eqnarray}
By using equation \ref{F in DF} we define a new parameter:
\begin{eqnarray}\label{Gamma}
& \Gamma^{\nu} = T^{a}\Gamma^{\nu a} =- T^{a}[\partial_{\mu}(\partial^{\mu}A^{\nu  a} - \partial^{\nu}A^{\mu  a}) ]
\nn & =T^{a}[ g \epsilon^{abc}\partial_{\mu}(A^{\mu b}A^{\nu c}) + g\epsilon^{abc}A^{b}_{\mu}(\partial^{\mu}A^{\nu  c} - \partial^{\nu}A^{\mu  c})
\nn & + g^{2}\epsilon^{abc}A^{b}_{\mu} \epsilon^{cdf}A^{\mu d}A^{\nu f} - J^{\nu a}]
\end{eqnarray}
It is easy to see that:
\begin{eqnarray}
\partial_{\nu}\Gamma^{\nu} =\partial_{\nu} \partial_{\mu}(\partial^{\mu}A^{\nu} - \partial^{\nu}A^{\mu}) = 0
\end{eqnarray}
So we can now construct a constant charge by defining (For more general definition see \cite{ABBOTT})
\begin{eqnarray}\label{Q def 1}
& Q = \int \Gamma^{0} \, d^{3}x
\nn & =  \int [ \partial_{\mu}(\partial^{0}A^{\mu} - \partial^{\mu}A^{0})] d^{3}x
\end{eqnarray}
Which one can find that it is constant by (see appendix \ref{secction:Non Abelian Charge definition} for the proof that $ Q $ is also gauge covariant):
\begin{align}
\partial_{0} Q = \int \partial_{0}\Gamma^{0} \,d^{3}x = \int \partial_{i}\Gamma^{i} d^{3}x = 0
\end{align}
 by using equation \ref{Gamma} $ Q $ can be define in equivalent form:
\begin{eqnarray}\label{Q def 2}
& Q = \int T^{a}[ g \epsilon^{abc}\partial_{\mu}(A^{\mu b}A^{0 c})
\nn & + g\epsilon^{abc}A^{b}_{\mu}(\partial^{\mu}A^{0  c} - \partial^{0}A^{\mu  c})
\nn & + g^{2}\epsilon^{abc}A^{b}_{\mu} \epsilon^{cdf}A^{\mu d}A^{0 f} - J^{0 a}]d^{3}x
\end{eqnarray}
where 
\begin{eqnarray}
& J^{\mu a} =iT^{a}[\phi^{+}D^{\mu}\phi - (D^{\mu}\phi^{+})\phi] = 
 \nn & iT^{a}[\phi^{+}\partial^{\mu}\phi - (\partial^{\mu}\phi^{+})\phi] + 2g\epsilon^{abc} \phi^{+}T^{b}A^{c\,\mu}\phi 
\end{eqnarray}
Now we trite $ Q $ as coupling constant, and the action of the system will be:
\begin{eqnarray}
& S = \int \lbrace D_{\mu}\phi^{+} D^{\mu}\phi - V(\phi^{+}, \phi ,Q)
\nn & - \frac{1}{4}F^{\mu\nu a}F_{\mu\nu}^{a} \rbrace d^{4}x
\end{eqnarray}
If we use the definition of equation \ref{Q def 1} or \ref{Q def 2} of $ Q $ then, by variation by $ A^{\mu} $ we get the original equation of motion plus delta term:
\begin{eqnarray}
& D_{\mu}F^{\mu\nu a} = J^{\nu a} 
\nn & + 2gT^{b}\epsilon^{abc} \phi^{+}T^{c}\phi \delta(t - t_{0}) 
\end{eqnarray}
variation on the action by $ \phi^{+} $ gives:
\begin{eqnarray}
& D_{\mu}D^{\mu}\phi + \frac{\partial V}{\partial \phi^{+}}
\nn & + (\int \frac{\partial V}{\partial Q} d^{4}x)[2i\partial^{0}\phi + 2gT^{a}\epsilon^{abc} T^{b}A^{c\,\mu}\phi ]\delta(t - t_{0})
\nn & +i\phi \partial^{0}\delta(t - t_{0}) = 0
\end{eqnarray}
The problematic term $ \phi \partial^{0}\delta(t - t_{0}) $ can not be transform away, so we need to conclude that:
\begin{eqnarray}
\phi(t = t_{0}) = 0
\end{eqnarray}

It is easy to see that the same constraint can be found directly from equation \ref{eq general constraint}.

\section{conclusion}
"The future physical theory should contain not only the basic
equations but also the initial conditions for them " [L.D. Landau according to I.M. Khalatnikov].
From this point we are motivated to construct a model where initial conditions can be found from the fundamental rules of physics, without the need to assume them, they will be derived.
In physics we deal with equations of motion that are obtained by varying the action with respect to different fields, here the question of the initial condition or boundary condition are normally separated from the equation of motion, and by giving them both we can solve the physical problem (like in many differential equation problems where the solution is determined by the initial condition).
Knowing just the equation of motion or just the initial conditions does not give the solution of the problem.
In this paper we showed that boundary condition can be contracted, or can be found by using the fact that If we have an action in the general form:
 \begin{equation}\label{eq:action with + g function conclution}
S=\int d^{4}x \{\mathcal{L} + \mathcal{G}(g (f))\}
\end{equation}
Where $ \mathcal{L} $ and $ \mathcal{G} $ are invariant under gauge transformation.
We reacquire that there is not exist any transformation which $ \mathcal{L} + \mathcal{G} \Rightarrow \mathcal{L} $.\\The function $ g(f(x)) $  have singular derivative on some surface $ f(x) = const $ , where $ f(x) $ is some analytic function.
The equation of motion will have the constraint equation: 
\begin{eqnarray}\label{eq general constraint conclution}
\frac{\partial^{2} \mathcal{G}}{\partial (\partial_{\mu}\phi)\partial g(x)}\partial_{\mu}f(x)\mid _{x\in f(x) = const} = 0
\end{eqnarray}
on the surface $ f(x) = const $ where $ g(x) $ have singular derivative.
Equation \ref{eq general constraint conclution} was proof.
Also we showed some example from the past and some new example of dynamical boundary condition or initial condition for non Abelian field.Those examples have been built around some physical philosophy, and the equation of motion produce the physical constraints. The MIT bag model has been built around the idea that the quarks are free to move in some cavity. The next examples show physical systems where the global charge in the universe are part of the local scalar field system, which is Mach like principle for global charge. There can be more examples. In the paper \cite{bernal} they contracted actions where the mass appears in the action that is, they put also a conserved quantity in the action. They showed that the Modified Newtonian Dynamics regime can be fully recovered as the weak-field limit of a particular theory of gravity formulated in the metric approach. They took Milgrom's acceleration constant as the fundamental quantity which couples to the theory. Since including the mass in the action affects also the equations of motion in particular at boundaries, we may get then boundary conditions, if the procedure is carried out consistently.
\section{acknowledgement}
I thank Prof Eduardo Guendelman for reviewing this article and for his useful comments.
I also like to thank him for advising me during the years of my Ph.D. 

\appendix
\section{Charge definition in terms of a dynamical field}
We will start by recalling the definition of area element of sub-manifold
\begin{equation}
x^{\mu}=\Phi^{\mu}(\lambda_{1},...,\lambda_{N})
\end{equation}
the element of area is:
\begin{equation}
d\tau^{\mu_{1},...,\mu_{N}}=\delta^{\mu_{1},...,\mu_{N}}_{\nu_{1},...,\nu_{N}}\frac{\partial\Phi^{\nu_{1}}}{\partial \lambda_{1}}....\frac{\partial\Phi^{\nu_{N}}}{\partial \lambda_{N}}d\lambda_{1}...d\lambda_{N}
\end{equation}
where:
\begin{eqnarray}
\delta^{\mu_{1},...,\mu_{N}}_{\nu_{1},...,\nu_{N}}=\begin{vmatrix} \delta^{\mu_{1}}_{\nu_{1}} & ... & \delta^{\mu_{N}}_{\nu_{1}} \\... &  &  \\ \delta^{\mu_{1}}_{\nu_{N}} &  & \delta^{\mu_{N}}_{\nu_{N}}  \end{vmatrix}
\end{eqnarray}
it can be also be written that:
\begin{equation}
d_{i}x^{\mu}=\frac{\partial \Phi^{\mu}}{\partial \lambda_{i}}d \lambda_{i}
\end{equation}
so the element of area is:
\begin{equation}
d\tau^{\mu_{1},...,\mu_{N}}=\delta^{\mu_{1},...,\mu_{N}}_{\nu_{1},...,\nu_{N}}d_{1}x^{\nu_{1}}...d_{N}x^{\nu_{N}}
\end{equation}
The dual element of the area of a 3-dimensional surface embedded in four dimension surface element is:
\begin{equation}
d\sigma_{\mu}=\frac{1}{3!}\epsilon_{\mu\nu\rho\sigma}d\tau^{\nu\rho\sigma} \\\,\, ,\,\, d\sigma=\frac{1}{4!}\epsilon_{\mu\nu\rho\sigma}d\tau^{\mu\nu\rho\sigma}
\end{equation}
where $ \epsilon_{\mu\nu\rho\sigma} $ is Levi Civita tensor where $ \epsilon^{\mu\nu\rho\sigma} $ is weight $ -1 $.
By the stokes theorem we have:
\begin{equation}
\oint{g^{\mu\nu}j_{\nu}\sqrt{-g} d\sigma_{\mu}}=\int{\partial_{\mu}(\sqrt{-g}j^{\mu})d\sigma}
\end{equation}
In our case $ j_{\mu}=\phi^{*}\stackrel{\leftrightarrow}{\partial}_{\mu}\phi-g'A_{\mu}\phi^{*}\phi $, if the current is conserved $ \partial_{\mu}(\sqrt{-g} j^{\mu})=0 $  so we have that:
\begin{equation}
\oint{j_{\nu}g^{\mu\nu}\sqrt{-g} d\sigma_{\mu}}=\int_{\mathcal{M}}{\partial_{\mu}(\sqrt{-g}j^{\mu}) d\sigma}=0
\end{equation}
So if we have close surface $ \Sigma=\Sigma_{1}+\Sigma_{2} $, where $ \mathcal{M} $ is the volume inside than we can have another conservation:
\begin{eqnarray}
& \oint_{\Sigma}{j_{\nu}g^{\mu\nu}\sqrt{-g} d\sigma_{\mu}}=
\nn &\int_{\Sigma_{1}}{j_{\nu}g^{\mu\nu}\sqrt{-g} d\sigma_{\mu}}-\int_{\Sigma_{2}}{j_{\nu}g^{\mu\nu}\sqrt{-g} d\sigma_{\mu}}=0
\end{eqnarray}
so:
\begin{equation}\label{bounders: theta def sigma}
Q \equiv \int_{\Sigma_{1}}{j_{\nu}g^{\mu\nu}\sqrt{-g} d\sigma_{\mu}}=\int_{\Sigma_{2}}{j_{\nu}g^{\mu\nu}\sqrt{-g} d\sigma_{\mu}}=const
\end{equation}
in the case $ d\sigma_{\mu} $ is space like, this represent the total amount of charge through the surface that entered over all times.\\
If we define theta function:
\begin{equation}
\theta(f(x) - f_{0})=\twopartdef {1}{f > f_{0}}{0}{f < f_{0} }
\end{equation}
where $ f(x^{\mu})=f_{0} $ on the surface $ \Sigma_{1} $, we also demand that $ \partial_{\mu}f(x)\neq 0 $ on the surface
then we have:
\begin{eqnarray}
& \delta^{\mu}(f(x) - f_{0})= \partial^{\mu}\theta(f(x) - f_{0}) = \frac{\partial\theta(f(x) - f_{0})}{\partial f(x)} \partial^{\mu}f(x) 
\nn & = \delta (f(x) - f_{0})\partial^{\mu}f(x)
\end{eqnarray}
So we can see equation \ref{bounders: theta def sigma} in anther way:
\begin{eqnarray}\label{Theta}
Q = \int_{\mathcal{M}_{1}}{(j^{\mu}\delta_{\mu}(f(x) - f_{0}))\sqrt{-g} d\sigma}
\end{eqnarray}
where the current define as $ j^{\mu} = \phi^{*}i\stackrel{\leftrightarrow}{\partial^{\mu}}\phi + g'A^{\mu}\phi^{*}\phi $


\section{Non Abelian Charge definition}\label{secction:Non Abelian Charge definition}
Now we will show that $ Q $ is also gauge invariant ($ \Gamma^{\nu} $ is not gauge invariant).
$ A^{\mu} $ transforms as:
\begin{eqnarray}\label{A transformation}
& A^{\mu} \rightarrow UA^{\mu}U^{-1} - \frac{i}{g}U\partial^{\mu}U^{-1}
\nn & = T^{a}Tr(U^{-1}T^{a}UA^{\mu}) - \frac{i}{g}T^{a}Tr(T^{a}U\partial^{\mu}U^{-1})
\end{eqnarray}
So $ \Gamma^{0} $ under the transformation of $ A^{\mu} $ transforms as (not gauge invariant):
\begin{eqnarray}\label{Gamma transforms}
&\Gamma^{0} \rightarrow  - \frac{i}{g} (\partial_{i}\partial^{0}U)\partial^{i}U^{-1} +\frac{i}{g} (\partial_{i}\partial^{i}U)\partial^{0}U^{-1}
\nn & - \frac{i}{g} (\partial^{0}U)\partial_{i}\partial^{i}U^{-1} +\frac{i}{g} (\partial^{i}U)\partial_{i}\partial^{0}U^{-1}
\nn & + T^{a}\lbrace Tr[U^{-1}T^{a}U(\partial_{i}\partial^{0} A^{i})] - Tr[U^{-1}T^{a}U(\partial_{i}\partial^{i} A^{0})] 
\nn & + Tr[(\partial_{i}\partial^{0}U^{-1}T^{a}U) A^{i}] - Tr[(\partial_{i}\partial^{i}U^{-1}T^{a}U) A^{0}] 
\nn & + Tr[(\partial^{0}U^{-1}T^{a}U)\partial_{i} A^{i}] -2 Tr[(\partial^{i}U^{-1}T^{a}U)\partial_{i} A^{0}] \nn &  + Tr[(\partial_{i}U^{-1}T^{a}U)\partial^{0} A^{i}]\rbrace
\end{eqnarray}
It is easy to see that:
\begin{eqnarray}\label{int by parts 0}
\int \frac{i}{g} (\partial_{i}\partial^{0}U)\partial^{i}U^{-1} \, d^{3}x =  - \int \frac{i}{g} (\partial^{0}U)\partial_{i}\partial^{i}U^{-1}\, d^{3}x 
\\ \int \frac{i}{g} (\partial_{i}\partial^{i}U)\partial^{0}U^{-1} \, d^{3}x = - \int \frac{i}{g} (\partial^{i}U)\partial_{i}\partial^{0}U^{-1} \, d^{3}x
\end{eqnarray}
and integration by parts leave just the surface terms of $ \int \Gamma^{0}d^{3}x $:
\begin{eqnarray}\label{int by parts 1}
& \int  T^{a}\lbrace Tr[U^{-1}T^{a}U(\partial_{i}\partial^{0} A^{i})] 
\nn & + Tr[(\partial_{i}\partial^{0}U^{-1}T^{a}U) A^{i}]
\nn & + Tr[(\partial^{0}U^{-1}T^{a}U)\partial_{i} A^{i}]
\nn &  + Tr[(\partial_{i}U^{-1}T^{a}U)\partial^{0} A^{i}]\rbrace d^{3}x = 
\nn &  Tr[U^{-1}T^{a}U \oint\partial^{0}A^{i}n_{i}dS] 
\nn & +  Tr[n_{i}A^{i}\oint\partial^{0}(U^{-1}T^{a}U)dS ]  =
\nn & Tr[U^{-1}T^{a}U \oint\partial^{0}A^{i}n_{i}dS]
\end{eqnarray}
Where we have used Stokes theorem and $ n_{i} $ is normal of surface and $ dS $ is integration of the surface.
We also delete the term $  Tr[n_{i}A^{i}\oint\partial^{0}(U^{-1}T^{a}U)dS ]  $ because the terms is eliminate on the surface.\\
We can also to do so for the other terms of \ref{Gamma transforms} under integration: 
\begin{eqnarray}\label{int by parts 2}
& \int T^{a}\lbrace  Tr[U^{-1}T^{a}U(\partial_{i}\partial^{i} A^{0})]
\nn & + Tr[(\partial_{i}\partial^{i}U^{-1}T^{a}U) A^{0}]
\nn & + 2 Tr[(\partial^{i}U^{-1}T^{a}U)\partial_{i} A^{0}] \rbrace d^{3}x = 
\nn &  Tr[U^{-1}T^{a}U \oint\partial^{i}A^{0}n_{i}dS] 
\nn & +  Tr[A^{0}\oint\partial^{i}(U^{-1}T^{a}U)n_{i}dS ]  =
\nn & Tr[U^{-1}T^{a}U \oint\partial^{i}A^{0}n_{i}dS]
\end{eqnarray}
So under transformation \ref{A transformation}, and the findings of equations \ref{int by parts 0},\ref{int by parts 1},\ref{int by parts 2} $ Q $ goes as:
\begin{eqnarray}
Q \rightarrow UQU^{-1}
\end{eqnarray}
Which means that $ Q $ is gauge covariant!
\bibliography{GeneralRulePrinciplebib2}
\bibliographystyle{plain}

\end{document}